# Intermetallic particle heterogeneity controls shear localization in high-strength nanostructured Al alloys

Tianjiao Lei [1], Esther C. Hessong [1], Jungho Shin [2,3], Daniel S. Gianola [2], Timothy J. Rupert [1,*]

[1] Department of Materials Science and Engineering, University of California, Irvine, CA 92697, USA

[2] Materials Department, University of California, Santa Barbara, CA 93106, USA

[3] Gangneung-Wonju National University, Gangneung-si, Gangwon-do, Republic of Korea

* Email: trupert@uci.edu

**Abstract**

The mechanical behavior of two nanocrystalline Al alloys, Al-Mg-Y and Al-Fe-Y, is investigated with in-situ micropillar compression testing. Both alloys were strengthened by a hierarchical microstructure including grain boundary segregation, nanometer-thick amorphous complexions, carbide nanorod precipitates with sizes of a few nanometers, and submicron-scale intermetallic particles. The maximum yield strength of the Al-Mg-Y system is measured to be 950 MPa, exceeding that of the Al-Fe-Y system (680 MPa), primarily due to a combination of more carbide nanorods and more amorphous complexions. Both alloys exhibited yield strengths much higher than those of commercial Al alloys, and therefore have great potential for structural applications. However, some micropillar specimens were observed to plastically soften through shear banding. Post-mortem investigation revealed that intermetallic-free deformation pathways of a few micrometers in length were responsible for this failure. Further characterization showed significant grain growth within the shear band. The coarsened grains maintained the same orientation with each other, pointing to grain boundary mechanisms for plastic flow, specifically grain rotation and/or grain boundary migration. The presence of intermetallic particles makes it difficult for both matrix and intermetallic grains to rotate into the same orientation due to the different lattice parameters and slip systems. Therefore, we are able to conclude that a uniform distribution of intermetallic particles with an average spacing less than the percolation length of shear localization can effectively prevent the maturation of shear bands, offering a design strategy for high-strength nanocrystalline Al alloys with both high strength and stable plastic flow.

# 1. Introduction

Achieving higher strength has been a long-standing research target for Al alloys in order to enhance their specific strength. For conventional coarse-grained Al alloys designed for high strength, the dominant strengthening mechanism is typically precipitation hardening, where a high number density of closely spaced precipitates form upon heat treatment and act as obstacles to dislocation movement. In Al 7075 alloys (primarily alloyed with Zn, Mg, and Cu), the precipitation process begins with Guinier-Preston (GP) zones, a metastable structure with a spherical morphology and sizes on the order of a few nanometers. The GP zones subsequently transition to a plate-like $\eta'$-$MgZn_2$ phase, which further evolves into equilibrium lath-shaped $\eta$-$MgZn_2$ precipitates (>50 nm in diameter) as aging time or temperature increases [1,2]. These precipitates significantly strengthen the materials as their contribution to the yield strength can be as high as 472 MPa [3]. For another Al alloy system [4], AA6111 (primarily alloyed with Mg, Si, Cu), the precipitation sequence also starts with the GP zones but with a needlelike morphology, and then transformation to a needle-shaped $\beta''$ phase ($Mg_5Si_6$) and a lath-like $Q'$ phase ($Al_4Cu_2Mg_8Si_7$) follows. An increase in the volume fraction of these precipitates from 0.2% to ~0.75% leads to an improvement of ~150 MPa in the yield strength. In addition to precipitation hardening, solid solution strengthening is another important mechanism in traditional Al alloys, as the presence of solute elements with a large lattice mismatch can retard dislocation motion and consequently strengthen the material. In an Al-8Ce-Mg (wt.%) alloy fabricated by high-pressure die casting [5], an addition of 0.75 wt.% Mg increased the lattice constant of Al from 4.0511 Å to 4.0540 Å, and this very small 0.07% misfit strain in turn enhanced the yield strength by 25%, from 92 MPa to 115 MPa.

Grain size refinement is another promising approach to improve yield strength because grain boundaries can serve as strong obstacles to dislocation motion. Zhao et al. [6] compared tensile yield strengths of 7075 Al alloys with different grain sizes, finding that the yield strength corresponding to an average grain size of 100 nm (550 MPa) was more than three times that of the coarse-grain counterpart (145 MPa). Grain boundary segregation can further increase yield strength due to the interaction between dopants and grain boundary plasticity mechanisms. For example, by using molecular dynamic simulations to study the deformation mechanism of nanocrystalline Al alloys, Babicheva et al. [7] predicted a tensile strength of 1.8 GPa for an Al-Co alloy with Co segregation to grain boundaries, while the strength of pure Al with the same average grain size was 1.4 GPa. The higher strength was attributed to a delay of grain boundary sliding and grain boundary migration due to the Co segregation. In an experimental study, Valiev et al. [8] observed that the strength of ultrafine-grained Al alloys prepared by high-pressure torsion exceeded the Hall-Petch scaling, which was possibly due to segregation of dopant atoms to grain boundaries that affected the emission and mobility of dislocations. Grain boundary segregation can also lead to structural transitions, such as the formation of amorphous grain boundary complexions [9]. Such amorphous complexions can have a positive strengthening effect on nanocrystalline materials, as Turlo and Rupert [10] showed that the complexions could act as strong dislocation pinning sites that increase the flow stress required for dislocation propagation, which is the rate-limiting mechanism of the plasticity for grain sizes between ~20 and 100 nm.

Although grain size refinement can give rise to exceptional strength, it can also lead to shear localization and catastrophic failure. Jia et al. [11] performed uniaxial compression tests on consolidated Fe with average grain sizes from tens of micrometers down to nanometers and observed that the deformation mode transformed from homogeneous to inhomogeneous with decreasing grain size, suggesting that shear banding becomes the dominant deformation mode when grain sizes are sufficiently small. These authors hypothesized that under an applied stress, larger grains would first undergo substantial plastic deformation while the surrounding small grains remained undeformed. When the stress was sufficiently high, small grains surrounding the larger grains would possibly rotate to orientations that were suitable for shearing, which triggered shear localization. In a computational study, Rupert [12] performed molecular dynamic simulations to study plastic strain distribution within nanocrystalline Ni and observed that the formation of shear localization could be either entirely through grain boundary deformation or through a combination of grain boundary sliding and grain boundary dislocation emission. In order to prevent strain localization, one effective approach is grain boundary engineering. Recently, Balbus et al. [13] used nanoindentation to investigate the mechanical behavior of nanocrystalline $Al_{85}Ni_{10}Ce_5$ (at.%) films, which showed shear offsets both under the tip and in the pileup regions in the as-deposited state. However, the shear localization was significantly suppressed after low-temperature annealing treatments. The suppression coincided with formation of amorphous complexions, suggesting that these complexions led to a preference for intragranular dislocation plasticity over grain boundary-mediated mechanisms and consequently a lower propensity for plastic localization.

In the present study, the mechanical behavior of two newly developed nanocrystalline Al alloys produced in a bulk cylinder form, Al-Mg-Y and Al-Fe-Y, with 2 at.% for each dopant element, was studied using in-situ scanning electron microscopy (SEM) micropillar compression testing. Both alloys contained a hierarchical microstructure consisting of grain boundary segregation, amorphous grain boundary complexions, nanorod precipitates, and larger intermetallic particles, all of which concurrently strengthen the material. The yield strength of the Al-Mg-Y system can be as high as 950 MPa, while the maximum yield strength of the Al-Fe-Y system is 680 MPa. The higher yield strength in the former alloy is mainly attributed to a higher number density of nanorod precipitates and Mg solutes within the matrix. Amorphous complexions may also strengthen the Al-Mg-Y alloy more effectively because of a possibly wider supercooled region of these complexions in this alloy. Although both alloys showed very high yield strengths, shear localization was also occasionally observed, albeit not in all samples tested. Post-mortem microscopy of the deformed pillars exhibiting localized deformation revealed pathways with a few micrometers in length that were free of intermetallic particles, the location of which were consistent with the dominant shear bands. Moreover, the grains within the shear bands significantly coarsened and exhibited the same orientation, pointing to grain rotation and/or grain boundary migration during the localized plastic flow. Therefore, we conclude that a uniform distribution of intermetallic particles with an average spacing much less than the percolation length of shear localization can effectively prevent the maturation of shear bands, due to a higher activation barrier for grain boundary mechanisms because of the dramatically different lattice parameters and slip systems between the matrix and intermetallic grains.

## 2. Materials and methods

### 2.1. Alloy fabrication

To synthesize bulk nanocrystalline alloy samples, powders of elemental Al (Alfa Aesar, 99.97%, -100+325 mesh), Mg (Alfa Aesar, 99.8%, -325 mesh) or Fe (Alfa Aesar, 99.9%, -20 mesh), and Y (Alfa Aesar, 99.6%, -40 mesh) were first ball milled for 10 h in a SPEX SamplePrep 8000M high-energy ball mill using a hardened steel vial and milling media. A ball-to-powder weight ratio of 10:1 was used with 3 wt.% stearic acid as a process control agent to prevent excessive cold welding. The milling process was conducted in a glovebox filled with Ar gas at an $O_2$ level <0.05 ppm to avoid oxidation. After milling, the alloyed powders were transferred into a ~14 mm inner diameter graphite die set, and then consolidated into cylindrical bulk pellets using an MTI Corporation OTF-1200X-VHP3 hot press consisting of a vertical tube furnace with a vacuum-sealed quartz tube and a hydraulic press. For the consolidation process, the powders were first cold pressed for 10 min under 100 MPa at room temperature to form a green body and then hot pressed for 1 h under 100 MPa at 585 °C, approximately equal to a homologous temperature ($T/T_m$) of 0.92 where $T_m$ = 663 °C is the melting temperature of pure Al [14]. The heating rate used to reach the target pressing temperature was 10 °C/min, and after hot pressing, the pellets were naturally cooled down to room temperature, which typically took more than 4 h. Readers are referred to Ref. [15] for more details on the consolidation process.

## 2.2. Microstructural characterization

The consolidated cylindrical pellets were first cut into half cylinders using a low-speed diamond saw. Subsequently, the cross-sectional surfaces were ground with SiC grinding paper down to 1200 grit and then polished with monocrystalline diamond pastes down to 0.25 μm prior to microstructural characterization. X-ray diffraction (XRD) measurements were conducted using a Rigaku Ultima III X-ray diffractometer with a Cu Kα radiation source operated at 40 kV and 30 mA and a one-dimensional D/teX Ultra detector. Phase identification and fraction were obtained using an integrated powder X-ray analysis software package (Rigaku PDXL). SEM imaging and backscattered electron (BSE) imaging were performed in an FEI Quanta 3D FEG dual-beam SEM/Focused Ion Beam (FIB) microscope. Scanning/transmission electron microscopy (S/TEM) paired with energy-dispersive spectroscopy (EDS) were used to examine the nanorod precipitate size and intermetallic chemistry inside of a JEOL JEM-2800 S/TEM, which was operated at 200 kV and equipped with a Gatan OneView IS camera and two dual dry solid-state 100 mm$^2$ EDS detectors. TEM-based orientation microscopy was performed using ASTAR$^{TM}$ (NanoMEGAS, Brussels, Belgium) hardware and software packages installed on the JEOL JEM-2800 S/TEM. The elemental distribution in the vicinity of the nanorod precipitates and grain boundaries was examined using high-angle annular dark field (HAADF)-STEM combined with EDS in a JEOL JEM-ARM300F Grand ARM TEM with double Cs correctors operated at 300 kV. For the HAADF imaging, a probe current of 35 pA together with an inner and outer collection angle of 106 and 180 mrad, respectively, were used. All TEM samples were fabricated using the FIB lift-out method [16] with a Ga$^+$ ion beam in the FEI Quanta 3D FEG dual-beam SEM/FIB microscope equipped with an OmniProbe. A final polish at 5 kV and 48 pA was used to minimize the ion beam damage to the TEM sample.

## 2.3. Micropillar compression testing

Micropillars were prepared in an FEI Quanta 3D Dual-Beam FIB/SEM using a FIB lathe milling method [17], which allows the final pillar to be taper-free to ensure a uniform stress state [18]. First, a Pt cap with a circular shape was deposited on the sample surface to protect the area of interest. Next, various milling procedures were carried out at 30kV. The first milling step was annular milling with a high ion beam current of 65 nA to remove material close to the area of interest so that a rough pillar shape was formed. The outer and inner diameters of the annular milling pattern were 70 μm and 30 μm, respectively, and the depth was 15 μm. Subsequently, a smaller annular milling (outer diameter of 35 μm, inner diameter of 12 μm) with a beam current of 30 nA was conducted to further remove extra material so that the shape of the pillar was a cylinder some taper angle, the diameter and height of which were approximately 12 μm. After the annular milling step, two rounds of lathe milling were performed. For the first round, a rectangular milling with a height of 13 μm and a width of 4 μm was performed every 20-degree rotation angle of the pillar to remove any pillar taper. The diameter of the pillar after this step was ~7 μm. Following the first round of lathe milling, a second lathe milling step was carried out by using a small beam current of 0.3 nA every 10-degree rotation angle of the pillar to polish the pillar surface and reduce the FIB damage due to the high beam currents used in previous steps. In order to make sure that the pillar deformation resembles bulk behavior, the final dimension of all pillars was ~5 μm in diameter and ~10 μm in height, which is much larger than the grain size of tens of nanometers [19]. Moreover, a height-to-diameter aspect ratio of ~2 was used to prevent plastic buckling [20].

The in-situ compression tests on the micropillars were performed using a FemtoTools nanomechanical testing system (Model FT-NMT03) under SEM observation. The load was applied by a flat platen with a cross section of 20 μm × 25 μm. The platen was milled from a flat Si MEMS-based micro-force sensor head (model FT-S200'200) with a ±200,000 μN force range and 0.5 μN force resolution. All tests were conducted in a displacement control mode using a subnanometer-resolution piezo-based actuation system. A nominal strain rate of ~$10^{-3}$ s$^{-1}$ was applied.

## 3. Results and Discussion

### 3.1. Undeformed Microstructure

The hierarchical microstructure was first characterized across all relevant length scales to allow relationships with mechanical behavior to be inferred. Figures 1(a) and (b) show XRD scans for the two systems, where extra peaks emerge in addition to the face-centered cubic (FCC) Al phase (squares), the position and intensity of which are consistent with $Al_3Y$ and $Al_{10}Fe_2Y$ for the Al-Mg-Y and Al-Fe-Y alloys, respectively. Consequently, one dominant intermetallic phase formed in each system, with the volume fraction of $Al_3Y$ being ~9% and that of $Al_{10}Fe_2Y$ being ~18%. The $Al_3Y$ has a trigonal structure and a space group of *R-3m* (166), with cell parameters of $a = b = 6.1950$ Å, $c = 21.1370$ Å, $\alpha = \beta = 90°$, and $\gamma = 120°$ [21]. The crystal structure of $Al_{10}Fe_2Y$ is orthorhombic and the space group is *Cmcm* (63), with cell parameters of $a = 8.9649$ Å, $b = 10.1568$ Å, $c = 9.0113$ Å, $\alpha = \beta = \gamma = 90°$ [22]. In order to study the spatial distribution of these intermetallic phases, BSE imaging was employed and revealed a relatively uniform distribution on the micrometer scale for both intermetallic phases (Figs. 1(b) and (e)). The average particle spacings for the Al-Mg-Y and Al-Fe-Y systems were estimated to be ~560 nm and ~280 nm, respectively. Most of the particles have submicron sizes, with a few larger exceptions on the order of a few micrometers. To further investigate the intermetallic particles at a finer scale, HAADF-STEM combined with EDS mapping was performed as shown in Figs. 1(c) and (f). These HAADF-STEM images further verified the larger particle spacing in the Al-Mg-Y alloy than the Al-Fe-Y alloy. However, at the nanometer scale, a spatial variation in the intermetallic spacing emerged as the particle density was higher in some areas than others, with the variation being much higher in Al-Mg-Y than in Al-Fe-Y, mainly due to the much smaller volume fraction of the intermetallic phase in the former alloy. In Al-Mg-Y, all intermetallic particles consisted of only Al and Y, verifying that the intermetallic phase was $Al_3Y$, while Mg atoms were uniformly distributed throughout the microstructure at this magnification. The preferential formation of an Al-Y intermetallic has been observed by Chen et al. [23] when Y was incorporated into Mg-Al alloys, and these authors reported that the Al-Y phase significantly strengthened the material. In the Al-Fe-Y system, most intermetallic particles were composed of all three elements, with one particle being the exception and containing Al and Y only (enclosed in a dashed oval in Fig. 1(f)), suggesting that $Al_3Y$ also formed in a small amount. However, the volume fraction of the $Al_3Y$ was much lower than that of the $Al_{10}Fe_2Y$, so $Al_{10}Fe_2Y$ was still the dominant intermetallic in the Al-Fe-Y system, as corroborated by the X-ray diffraction phase analysis.

TEM was subsequently performed to study nanoscale features comprising the hierarchical microstructure, with Figs. 2(a) and (b) showing representative bright-field (BF) TEM micrographs for the two alloy systems. All grains have an equiaxed shape and a relatively uniform size well below 100 nm, with the average TEM grain size of Al-Mg-Y ($58 \pm 19$ nm) being slightly larger than that of Al-Fe-Y ($54 \pm 17$ nm). In addition, plenty of precipitates with a rod shape (termed "nanorods") were observed at grain boundaries (indicated by yellow arrows) with sizes of a few nanometers wide and tens of nanometers long. To study the structure of the nanorods, high-resolution HAADF-STEM was used, which revealed the atomistic details of the nanorod interior (last two panels in Figs 2(a) and (b) are micrographs and corresponding Fourier-filtered images). Our previous work [15] showed that the interior of the nanorods consisted of Al and C. Consequently, the bright spots in the Fourier-filtered images most likely correspond to Al, since its atomic weight is larger than that of C. The atomic arrangement of the Al atoms matches that of the $Al_4C_3$ phase, one schematic illustration of which is also presented as the inset. Therefore, these nanorods are assigned as aluminum carbides, consistent with those determined for a different Al-rich alloy system in our earlier work [24]. Since the nanorods possess an elongated morphology, the length and width were both measured and the corresponding cumulative distribution functions are plotted in Fig. 2(c). For Al-Mg-Y, the nanorods are $23.6 \pm 12.6$ nm long and $5.5 \pm 1.2$ nm wide, while those in Al-Fe-Y have an average length of $21.0 \pm 8.9$ nm and width of $4.8 \pm 1.1$ nm.

Therefore, the nanorods in Al-Mg-Y are slightly larger than in Al-Fe-Y, consistent with the trend in matrix grain size. In fact, the ratio of the grain size between the two alloys (1.07) is very close to that of the nanorod size (1.12 for length and 1.15 for width), suggesting that matrix grains and nanorods share similar growth kinetics, possibly due to the amorphous grain boundary complexions serving as both nucleation sites and reservoirs for solute atoms to the nanorods during subsequent growth [24]. Rod-shaped precipitates are often key strengthening components in Al and Mg alloys [25,26,27], with various factors affecting the strengthening effects including orientation, number density, and aspect ratio [25]. In the present study, no specific orientation between the nanorods and matrix phases was observed, and the nanorod number densities are ~18000/$\mu m^3$ and ~12000/$\mu m^3$ for Al-Mg-Y and Al-Fe-Y, respectively. For the length-to-width aspect ratio, both alloy systems exhibit values that are close to 4.3. Therefore, the strengthening effect due to the nanorods is expected to be higher in Al-Mg-Y than in Al-Fe-Y because of the higher number density in the former alloy.

Figures 3(a) and (d) present HAADF-STEM micrographs of the two systems, where grain boundaries and nanorod edges appear brighter than the matrix and nanorod interior, suggesting an enrichment of elements that are heavier than Al. Since the atomic weight of Mg is smaller than that of Al, the brighter contrast in Al-Mg-Y comes from Y, while brighter contrast from both Y and Fe is possible in Al-Fe-Y. Figures 3(b) and (c) show elemental mapping of representative examples of a grain boundary and a nanorod in Al-Mg-Y, which confirms the segregation of Y to both the grain boundary and nanorod edge. In addition, the concentration of Mg is higher at those regions, pointing to co-segregation of Mg and Y. Since stearic acid ($C_{17}H_{35}CO_2H$) was added during ball milling to prevent cold welding, the distribution of C atoms is also shown. No enrichment of C at the grain boundaries is observed, but the nanorod interior clearly contains a large amount of C atoms, which further verifies the phase to be $Al_4C_3$. For Al-Fe-Y, co-segregation of Fe and Y is also observed at both grain boundaries (Fig. 3(e)) and nanorod edges (Fig. 3(f)), and the nanorod interior is composed of Al and C. The co-segregation in both alloy systems should significantly contribute to their exceptional thermal stability, as dopants at grain boundaries can effectively stabilize nanosized grains owing to a decreasing drive force for grain growth and/or a pinning effect acting to suppress grain boundary migration [28,29]. Similarly, the co-segregation at the nanorod edges may also help stabilize the nanorods against rampant coarsening. From the EDS mapping of nanorods in both alloys (Figs. 3(c) and (f)), the co-segregation along the longer sides is more pronounced than that along the shorter edges, especially for Y atoms. Consequently, the stabilization of the longer side is likely to be stronger, leading to a faster growth along the nanorod length and therefore an increasing length-to-width aspect ratio as microstructure evolves further [24].

Since the segregation of dopant elements to grain boundaries may give rise to structural transitions (e.g., formation of amorphous grain boundary complexions), the structure of the grain boundaries was also examined. Amorphous complexions were observed in both systems, with Fig. 4 showing high-resolution TEM micrographs of representative amorphous complexions (enclosed in dashed lines). The complexion thickness is similar for both alloys, ~2-3 nm, which is also close to those in a naturally cooled nanocrystalline Al-Ni-Y system that was hot pressed at the same temperature [15], suggesting that the segregation of Y may play a more important role in the complexion formation than that of the transition metal elements. Although only one image of an amorphous complexion is presented for each alloy to avoid recreating information shown in prior work [15], these features are expected to widely formed in the microstructure due to the large atomic size mismatch and negative pair-wise mixing enthalpy values, both of which are beneficial for the amorphous complexion formation [30]. Previous studies [13,31] showed the existence of amorphous complexions in a similar alloy, Al-Ni-Ce, using both nanobeam diffraction and synchrotron X-ray scattering experiments. Halo rings and diffuse features emerged from the diffraction and scattering patterns, respectively, both of which point to the amorphous regions. In addition, more than 50 amorphous complexions were directly characterized using HRTEM for another nanocrystalline ternary system, Cu-Zr-Hf, that was designed with the same materials selection criteria [32]. It is worth mentioning that the retention of the amorphous complexions after a very slow cooling rate points to the outstanding stability of these features in the two alloys, as slow cooling can lead to transitions back to the ordered boundary state. The processability of these alloys as a direct consequence is therefore good, as the mechanical properties to be reported in the next section are obtained without additional annealing and/or quenching steps.

## 3.2. Mechanical Behavior

The mechanical behavior of the two alloys was studied using in-situ micropillar compression testing. Three representative pillars for each alloy system are presented in Fig. 5, where two Pt fiducial markers were deposited on each end of the gauge section to enable a more accurate calculation of the specimen strain. Intermetallic particles are easily discerned because of their different contrast from the matrix, most likely due to slight preferential milling of the matrix because the intermetallic phases are much harder than the Al phase [33].

Figure 6 shows the engineering stress-strain curves from the micropillar compression experiments. Most pillars exhibited a stress-strain response with very similar slopes within the elastic region, pointing to a consistent alignment between the sensor head and pillar. For the Al-Mg-Y system (Fig. 6(a)), five pillars were examined and demonstrated a diversity of behavior;

nevertheless, our following analysis shows that the behavior can be classified into two categories: (1) stable plastic flow (red curves) and (2) strain localization into shear bands (green curves), which will be discussed below through post-test examination. For Pillars 1 (circles) and 2 (up-pointing triangles), the measured yield strengths were the highest (950 MPa and 890 MPa, respectively), and the stresses decreased rapidly after yielding. The measured yield strengths of Pillars 3 (squares) and 4 (down-pointing triangles) were the lowest (500-600 MPa). After yielding, the stresses decreased at a lower rate than those for Pillars 1 and 2. It should be noted that the slope of the elastic region for Pillar 3 was smaller than that for all other pillars, possibly due to porosity within the pillar and/or in the material underneath the pillar. For Pillar 5 (diamonds), the measured yield strength was lower than the highest values (~680 MPa), and the stress seemingly increased after yielding, suggestive of strain hardening. We show below that this is not the case and instead an artifact of the strain localization failure mode. For the Al-Fe-Y alloy (Fig. 6(b)), six pillars were studied and all exhibited repeatable yield strengths with an average value of $630 \pm 44$ MPa. After yielding, two types of behavior were identified – one showed increasing stress and the other demonstrated the opposite trend. Similar as the Al-Mg-Y alloy, the increasing stress is not a real strain hardening effect, as will be shown in the following section.

### 3.2.1. Deformation modes – Shear localization versus stable plastic flow

To investigate the different categories of deformation behavior, each pillar was examined after the compression tests. Figure 7 presents four deformed pillars of Al-Mg-Y with two SEM images taken from different rotation (R) and tilt (T) angles for each. Pillar 1 (Fig. 7(a)) exhibited the highest strength and a rapid strain softening after yielding. The corresponding front view image shows that small cracks formed across apparent intermetallic particles at the bottom of the pillar (more clearly presented in the zoom-in view shown as an inset). These cracks likely caused the apparent softening in the corresponding stress-strain curves at larger applied strains, due to a dramatic decrease in the stress-carrying capability of the material. Due to their brittle nature, intermetallics are often preferential crack initiation sites and provide crack propagation paths in multiphase alloys [34]. When the microstructure is anisotropic, the direction of the applied load affects the intermetallic particle cracking process. For example, Agarwal et al. [35] studied the cracking of Fe-rich intermetallic particles in an extruded 6061 Al alloy (with grain size >> particle size) by performing room-temperature compressive testing, and observed a difference in the number fraction of cracked particles with different loading orientations with respect to the extrusion direction, due to the anisotropic microstructure and particle rotation during deformation. In the present study, the grains have an equiaxed shape and are much smaller than the intermetallic particles, and therefore the particle cracking process is most likely independent of the loading direction. The yield strengths of Pillars 3 and 4 were much lower than the others, and the corresponding micrographs (Figs. 7(b) and (c)) reveal the formation of dominant shear bands. In Pillar 3 (Fig. 7(b)), the shear band crossed the middle region of the pillar, while the localization traversed from the middle left to the bottom right in Pillar 4 (Fig. 7(c)). Pillar 5 (Fig. 7(d)) also experienced shear localization as the top region clearly sheared downwards. However, this localized deformation resulted in an increased contact area at the top, and consequently, an apparent increasing stress after yielding for this pillar. The post-mortem inspection indeed confirms that this is not a hardening effect but rather a geometrical artifact due to the increasing top area. The true final stress, calculated using the final top area, is also shown in Fig. 6(a). This value is clearly much lower than the original engineering stress and even below the yield strength, suggesting softening in reality for this sample as well. Therefore, the Al-Mg-Y system exhibited two deformation modes – one is stable plastic flow (Pillars 1 and 2) and the other is shear localization (Pillars 3, 4, and 5). For those which experienced stable plastic flow, high yield stresses were observed with the corresponding strengthening mechanisms to be discussed in detail in the next section. However, if shear localization occurred, the yield stresses were observed to be much lower, and more scattered due to an inherent stochastic nature of the shear banding determined by the intermetallics distribution.

Figure 8 shows deformed pillars for the Al-Fe-Y alloy. For Pillars 2 and 4 (Figs. 8(a) and (b)), their stresses decreased after yielding, most likely due to crack formation and propagation as several cracks were observed at the bottom of the Pillar 4. For Pillars 3 and 5 (Figs. 8(c) and (d)), the stress-strain curves showed increasing flow stress after yielding. However, the images of the deformed pillars again clearly reveal an increased contact area due to the top region shearing downwards. Therefore, the deformation modes observed in Al-Fe-Y can be assigned to those observed in Al-Mg-Y, including localized deformation within shear bands and stable plastic flow.

All deformed pillars were FIB cross-sectioned at the mid-plane so that a more in-depth examination of the microstructure within the failed pillars could be examined. Figure 9 shows secondary electron SEM micrographs for pillars that experienced shear localization. Figures 9(a)-(c) correspond to the Al-Mg-Y alloy while Figs. 9(d) and (e) are for the Al-Fe-Y system. All images reveal darker pathways (marked by dashed lines) which aligned with the location of the shear bands. These pathways are dark because they are free of intermetallic particles, as more clearly presented in the magnified images (bottom row). For the Al-Mg-Y system, the width of the pathway in Pillar 3 is the largest (~1 μm wide), while the intermetallic-free pathways in Pillars 4 and 5 are thinner (only a few hundred of nanometers wide). The propagation lengths of all the pathways are on the order of a few micrometers, much larger than the average intermetallic particle spacing (~560 nm) in the Al-Mg-Y system if the particles are uniformaly distributed. It is worth noting that the pathway in Pillar 5 localized just above one large intermetallic particle that appears bright on the right side of the pillar, suggesting that the propagation of the shear localization circumvented the intermetallic phase. In the Al-Fe-Y system, Pillars 3 and 5 experienced shearing of the top area, and darker pathways free of intermetallic particles were observed at the top as well with widths of ~100-200 nm and lengths of a few micrometers. For Pillar 5 (Fig. 9(e)), the pathway also went just above one large intermetallic particle in the middle of the pillar. Therefore, both the spatial and size distributions of intermetallic particles seem to affect the propagation of the localized deformation, as the percolation of the shear bands requires lengths much larger than the average particle spacing of uniformly distriubted intermetallics and the shear bands deflect away from prominent intermetallic particles.

Cross-sectional images of the pillars demonstrating stable plastic flow (without strain localization into dominant shear bands) are presented in Fig. 10. Unlike the pillars that experienced shear localization, no darker intermetallic-free pathways were observed in these pillars. Rather, the bright spots associated with the intermetallics were distributed throughout the specimen. Since the volume fraction of the intermetallic phase in Al-Mg-Y is lower than in Al-Fe-Y, the chances of the shear band propagation should be higher in the former system and we indeed observe more prominent strain localization in the Al-Mg-Y alloy. In addition, shear banding can occur in any region due to the lower density of intermetalic particles, which results in a variation in yield strength as shown in Fig. 6(a). However, when the shear localization is avoided, the Al-Mg-Y system is intrinsically stronger as its flow stress can reach up to ~1 GPa, suggesting more potent microstructural strengthening features. It is worth noting that the shear localization behavior for the present alloys may depend more on the microstructure than the sample size, as the percolation length of the shear bands at maturation is much larger than the average intermetallic spacing in the micropillars. Consequently, for bulk-size samples, shear localization may also occur if an intermetallic-free pathway on the similar length scale as the percolation length of shear bands exists.

HAADF-STEM of a deformed Al-Mg-Y pillar showing pronounced localization (Pillar 4) was employed to verify the internal distribution of the intermetallic phase, as shown in Fig. 11 along with the SEM micrographs to demonstrate the location of the shear band (enclosed in yellow dashed lines in each image). The shear band is about 200 nm wide and no intermetallic particles were observed in the HAADF images, further confirming that the shear band traverses a region free of intermetallic particles. The effect of secondary phases on shear banding has been widely studied in metallic glasses since these materials often fail catastrophically through strain localization within dominant shear bands. For instance, by forming ductile β phase dendrites ($Zr_{71}Ti_{16.3}Nb_{10}Cu_{1.8}Ni_{0.9}$, in at.%) within a Zr-based (Zr-Ti-Nb-Be-Cu-Ni) metallic glass during cooling from the melt, Hays et al. [36] observed that the propagation of individual shear bands was confined to regions with sizes comparable to the dendrite dimension. The dendritic β phase had a body-centered cubic structure and was uniformly distributed within the matrix with a volume fraction of ~25%. Therefore, the dendritic β phase was suggested to serve as both heterogeneous nucleation sites for shear bands and pinning points on the shear band propagation. In addition to metallic glasses, dispersed dendritic phases with sizes of a few micrometers have been observed to effectively prevent shear band propagation in Ta-rich nanostructured alloys with grain sizes below 50 nm [ 37 ]. The investigated Ta alloys included $Ti_{60}Cu_{14}Ni_{12}Sn_4Ta_{10}$ and $Ti_{60}Cu_{14}Ni_{12}Sn_4Nb_{10}$, while the dendritic phases were identified to be body-centered cubic Ti(Ta,Sn) and Ti(Nb,Sn), respectively. Shear bands were observed to bypass or stop at dendrites, indicating that the dispersed dendrite network obstructed highly localized shear banding and consequently prevented shearing-off through the whole sample. However, due to the ductile nature of the dendritic phase, shear bands would occasionally cut through dendrites on a few occasions in that study. In contrast, in the present study, no shear bands were found to cut through any intermetallic particles, indicating that the intermetallic phases are hard and brittle obstacles.

The grain morphology and size in the vicinity of the shear band was also examined using BF-TEM and is shown in Fig. 12(a). It is clear that grains within and close to the shear band have a much larger size (>200 nm) than those far away from the localized deformation (~60 nm, shown in micrograph outlined in green), pointing to targeted grain coarsening within the sheared region. Because of the increased grain size, dislocations were observed in the grain interiors, with one example presented in the magnified image outlined in red. Compared to nanosized grains, where intragranular dislocations are rapidly absorbed in the grain boundaries, these larger grains favor intragranular dislocation accumulation because more dislocation sources can be found in a single grain [38].

Shear localization in nanocrystalline metals and alloys has been attributed to grain-boundary-based mechanisms, including grain rotation [39] and grain boundary migration [40]. For the grain rotation mechanism, neighboring nanosized grains rotate into a similar orientation in order to reduce the barrier between them [41], while in the grain boundary migration scenario, grain boundaries move by atomic shuffling and free-volume migration to relieve the stress built up across neighboring grains [42]. Both cases result in grain coalescence along the shear direction, leading to larger grains within the shear band. In order to examine the grain orientations of this deformed region, ASTAR automated crystal orientation mapping was performed on both the deformed pillar with shear localization and an undeformed sample, as shown in Fig. 12(b). The shear band clearly had a preferred crystallographic texture as all grains within it have the same orientation, while the grains in the undeformed condition were randomly orientated. Such shear banding-induced grain growth has been observed in other nanocrystalline alloy systems as well. For example, Khalajhedayati and Rupert [43] employed both micropillar compression and nanoindentation techniques to study localized deformation in a nanocrystalline Ni-W (initial average grain size of 5 nm), and observed obvious grain coarsening and texturing within intense shear localization. In a large-scale atomistic simulation study on sliding experiments of nanocrystalline Fe, Romero et al. [44] demonstrated that extensive grain coarsening through grain boundary migration and simultaneous lattice rotation occurred until an optimal plastic slip orientation aligned with the sliding direction, and then subsequent sliding was accommodated by localized shear bands.

To verify the shear banding-induced grain growth mechanism, the microstructure of another pillar that experienced shear localization was examined. Figures 13(a) and (b) are low-magnification BF-STEM and enlarged BF-TEM micrographs from the deformed Pillar 3 of Al-Mg-Y, which exhibited dominant shear bands across the middle region of the pillar (Fig. 7(b)). The low-magnification BF-STEM micrograph presents an overview of the entire microstructure and clearly shows that significant grain coarsening occurred in the middle region, coinciding with the location of the dominant shear band. For regions outside the shear band, Fig. 13(b) shows that all grains remain nanosized, and that the grain structure is identical to that far away from the dominant shear band in the deformed Pillar 4 (outlined by green lines in Fig. 12(a)). Therefore, pillars experiencing shear localization only exhibit grain coarsening within the dominant shear bands, due to the activation of grain-boundary-mediated mechanisms. For pillars that underwent stable plastic flow, the microstructure of the deformed Pillar 1 of Al-Mg-Y can be used as a representative example. Figure 13(c) presents a low-magnification BF-STEM micrograph along with a zoomed-in view of one selected region, where only nanocrystalline structure was observed. Figure 13(d) is a HAADF-STEM image showing the size, morphology, and distribution of the nanorod carbides in the deformed Pillar 1, all of which are consistent with those observed in Fig. 3(a), showing the undeformed state of the sample. The microstructure after the stable plastic flow was further examined by BF-TEM, as shown in Fig. 13(e), which is very similar to those away from localized deformation in the deformed Pillars 3 and 4, with no coarsening. Consequently, there is no evidence that any significant microstructural variation occurred during deformation for the pillars that experienced stable plastic flow.

One effective approach that has been shown to prevent localized deformation into shear bands for nanocrystalline alloys is to form amorphous grain boundary complexions, since these complexions can lead to a preference of intragranular dislocation plasticity over grain boundary dominated deformation mechanism. Balbus et al. [13] performed nanoindentation tests on a nanocrystalline Al-Ni-Ce alloy and observed a transition from strain localization to homogeneous deformation with increasing annealing temperature, which coincided with the formation of amorphous complexions. In the present study, amorphous complexions were also observed (Fig. 5) while a few pillars exhibited shear localization, suggesting that other microstructural features, i.e., intermetallic phases, can still significantly affect shear banding. Due to the dramatically different lattice parameters between the intermetallic and matrix, we believe that it will be more difficult for embryonic shear bands to evolve to maturity if intermetallic particles are uniformly distributed within the matrix. As a result, grain coalescence can be effectively prevented, and the probability of strain localization will be low. When the particles were heterogeneously distributed in the matrix, the distribution can be viewed as a combination of loosely packed (with a larger inter-particle spacing) and closely packed regions (with a smaller inter-particle spacing). Chen et al. [45] performed compression tests on a Mo particle reinforced Mg-based bulk metallic glass composite with an average inter-particle spacing of ~38 μm, and observed that it showed an amount of large plasticity (7% strain-to-failure) without the propagation of long-range shear bands. When the inter-particle spacing was dramatically increased, e.g., to a value of ~90 μm, the strain-to-failure was only 1% [46]. This improved plasticity corresponding to the smaller inter-particle spacing was attributed to a stronger restriction of the shear-band extension, because the harder Mo particles can retard the consecutive shear-band propagation along the principal shear plane by absorbing partial shear stresses. Hofmann et al. [47] further proposed a criterion for limiting the shear band extension within metallic glass composites, which is that the microstructural length scales, e.g., the size of secondary phases and the spacing between the secondary phases, need to be comparable with a characteristic length scale that is associated with the maximum spatial extension of shear bands. Despite the fact that we are working with nanocrystalline alloys rather than metallic glass composites, these finding are consistent with the present study. We observed that intermetallic particles distributed heterogeneously to give regions with large particle spacings of a few micrometers result in dominant shear bands with propagation lengths of a few micrometers. In contrast, when the intermetallics were distributed relatively uniformly with an average spacing of a few hundred of nanometers, no such dominant shear bands formed. Taken as a whole, we hypothesize that a uniform distribution of intermetallic particles with an average spacing much less than the percolation length of shear localization can effectively prevent the maturation of dominant shear bands in nanocrystalline materials. On the other hand, if the spatial heterogeneity of the intermetallic particles approaches the length scale required for shear band maturity, shear localization will be rampant.

### 3.2.2. Strengthening mechanisms

The micropillar compression testing revealed that both alloys exhibited yield strengths much higher than those of commercially available Al alloys. For instance, the tensile yield strength of one commercial Al 7075 alloy, one class of the highest strength Al alloys available and often used in transportation and aerospace applications, ranges from 145 MPa to 476 MPa, depending on the temper treatment [48]. Another popular high-strength Al alloy that is also often used in the aerospace industry, Al 2024, can exhibit tensile yield strengths from 324 MPa to 400 MPa under various different heat treatments [49]. Although the present study reports only compressive yield strengths, these values can be approximated as the tensile counterparts since no obvious asymmetry in the tensile/compressive yield strengths was observed for a multi-phase nanocrystalline Al alloy [50]. When no shear localization occurred, Al-Mg-Y can reach a higher maximum yield strength (950 MPa) than Al-Fe-Y (680 MPa), consistent with the trend observed in our prior nanoindentation experiments [15]. Table I lists the hardness values from the prior nanoindentation experiments as well as the yield strengths of the pillars without shear localization, where Al-Mg-Y is clearly stronger than Al-Fe-Y by both metrics. Moreover, the ratio of hardness to yield stress is approximately three for both, which follows Tabor's relation [51] of $H = C \cdot \sigma$, where $H$ is the hardness, $\sigma$ is the yield stress, and $C$ is the constraint factor (often equal to 3 for metallic materials [52]). The high strengths of these alloys are mainly attributed to a large volume fraction of grain boundaries since the grain sizes are only ~50 nm. According to the Hall-Petch relation [53,54], the yield stress improvement due to grain boundary strengthening, $\sigma_{GB}$, can be expressed as:

$$\sigma_{GB} = \sigma_0 + kd^{-1/2} \tag{1}$$

where $\sigma_0$ represents the friction stress for individual dislocations (10 MPa for pure Al [55]), $k$ is a constant (0.08 MPa·m$^{-1/2}$ for pure nanocrystalline Al [56]), and $d$ is the average grain size. Therefore, for pure nanocrystalline Al system, $d$ = 58 nm (average grain size of Al-Mg-Y) and $d$ = 54 nm (average grain size of Al-Fe-Y) will give rise to $\sigma_{GB}$ values of 342 MPa and 354 MPa, respectively. Grain boundary segregation was observed in both systems (Fig. 3) and should affect the strength as well. Vo et al. [57] investigated the effect of grain boundary segregation on the yield strength of dilute nanocrystalline Cu systems doped with Nb, Ag, or Fe using molecular dynamics, where all of the samples had the same average grain size. These authors observed that the dopant elements that lowered the grain boundary energy could dramatically increase yield strength, suggesting that the grain size was not the only factor affecting strength. Therefore, a model including both the grain size and grain boundary energy contribution was proposed as [57, 58],

$$\frac{1}{\sigma} = \frac{1}{k_1+k_2/d^{1/2}}\left(1 - \frac{N_{GB}}{N_{total}}\right) + k_3 E_{GB}\left(\frac{N_{GB}}{N_{total}}\right) \qquad (2)$$

where $\sigma$ is the flow stress, $d$ is the grain size, $k_1$, $k_2$, and $k_3$ are fitting parameters, $N_{GB}$ and $N_{total}$ represent the number of atoms in grain boundaries and the total number of atoms, respectively, and $E_{GB}$ is the specific grain boundary energy and is defined as the excess grain boundary energy per grain boundary atom. The first term on the right-hand side of Eqn. (2) corresponds to traditional Hall-Petch strengthening and has a similar form as Eqn. (1), while the second term represents grain boundary sliding, where dopant elements that can decrease $E_{GB}$ will increase the yield strength. The ratio of the number of atoms in grain boundaries to that in the whole sample ($N_{GB}/N_{total}$) will determine the relative contribution of each term to the total strength. Since the parameters $k_3$, $E_{GB}$, and $N_{GB}/N_{total}$ are unknown for the present two nanocrystalline Al alloys, the exact values of Eqn. (2) cannot be obtained at this time. However, Ref. [57] showed that for the Cu-Nb system, the yield strength first increased linearly with increasing dopant concentration up to 1.2 at.% and then gradually approached saturation at higher concentration, with the relative yield stress increase due to the 1.2 at.% dopants being ~60%. If we assume that grain boundary segregation can also enhance the yield stresses of the present alloys by ~60%, the combined contribution of both grain size and grain boundary energy to the yield stress will be 547 MPa and 566 MPa for Al-Mg-Y and Al-Fe-Y, respectively.

The second strengthening effect comes from the carbide nanorods at grain boundaries, as a large number density of the nanorods were observed in both systems (e.g., Figs. 4(a) and (d)). For precipitates formed in conventional alloys, contributions to the material strength come either through resistance to dislocation shearing or an Orowan dislocation bypassing mechanism, since the precipitates often nucleate and grow in the grain interior and exhibit much smaller size than the matrix grains. However, in the present alloys, the nanorod precipitates are located at grain boundaries, so neither of the above-mentioned mechanisms will apply. Previous studies have shown that precipitates at grain boundaries contribute to the material strength through a mechanism termed grain boundary precipitate strengthening [59,60,61]. Zhang et al. [60] investigated the effect of grain boundary precipitates on creep deformation of Fe-15Cr-25Ni (wt.%) alloys with two carbon concentrations (0.002 wt.% and 0.086 wt.%). These authors observed that the sample with the higher carbon content exhibited creep behavior similar to that of dispersion hardening alloys, which was attributed to carbides formed at grain boundaries. TEM micrographs revealed a much higher dislocation density at grain boundaries in the high-carbon alloy, pointing to a strong obstruction of the intergranular carbides to dislocation motion. Zhang et al. developed a mechanical model to estimate the obstacle stress due to the boundary carbides, which can be expressed as:

$$\sigma_{obs} = m(2bKG/d)^{\frac{1}{2}}\sigma_{app}^{\frac{1}{2}} \tag{3}$$

where $m$ (ranging from 0 to 1) is a stress concentration factor taking account of intergranular particle density [61], $b$ is the Burgers vector, K is a constant and often taken as 20 for metals and alloys [62], $d$ is the grain diameter, and $\sigma_{app}$ represents the applied stress. Based on our TEM measurements, there are roughly ~1.53 and ~0.94 nanorods per matrix grain for Al-Mg-Y and Al-Fe-Y, respectively. Therefore, we account for three nanorods at the edges of each grain in Al-Mg-Y and two for Al-Fe-Y, so that the effective numbers of nanorods per grain are 1.5 and 1 for Al-Mg-Y and Al-Fe-Y, respectively, since each nanorod at a boundary is shared by two adjacent grains. If we assume that the grains have a hexagonal shape with six edges, then $m$ can be approximately as 1/2 for Al-Mg-Y and 1/3 for Al-Fe-Y. Plugging in the average grain size value of 58 nm and 54 nm for Al-Mg-Y and Al-Fe-Y, respectively, and assuming $\sigma_{app}$ is equal to the average yield stress corresponding to the stable flow condition for each alloy, the obtained $\sigma_{obs}$ values are ~1100 MPa and ~620 MPa for Al-Mg-Y and Al-Fe-Y, respectively. However, these two values dramatically overestimate this effect because the entire body of the nanorod carbides are not located at grain boundaries. Furthermore, Eqn. (3) is based on a matrix grain size of 240 μm [60], which is much larger than that in the present study (~50 nm), and modifications may be necessary for nanocrystalline alloys.

Because the intermetallic phases are harder than the Al matrix in both alloys, they are expected to have a positive effect on strengthening the material. One way to estimate their contribution is by employing a rule-of-mixture model [63],

$$\sigma = \sigma_m(1-f) + \sigma_{im}f, \tag{4}$$

where $\sigma$, $\sigma_m$, and $\sigma_{im}$ are the yield stress of the overall material, the matrix, and the intermetallic phase, respectively, and $f$ is the volume fraction of the intermetallic particles. The volume fractions of Al$_3$Y and Al$_{10}$Fe$_2$Y obtained from XRD experiments were ~9% and ~18%, respectively. However, due to the lack of available data on the yield stresses of the two intermetallic phases, the exact values of the yield stress improvement from this effect cannot be obtained rigorously. If we leave the strength as an unknown variable by taking $\sigma_{im}^{Al_3Y} = C_1 * \sigma_m$ and $\sigma_{im}^{Al_{10}Fe_2Y} = C_2 * \sigma_m$, where $\sigma_{im}^{Al_3Y}$ and $\sigma_{im}^{Al_{10}Fe_2Y}$ represent the yield stress of Al$_3$Y and Al$_{10}$Fe$_2$Y, respectively, and $C_1$ and $C_2$ are constants, then the contribution due to the intermetallic phases, defined as $\sigma - \sigma_m$, are $0.09(C_1 - 1)\sigma_m$ and $0.18(C_2 - 1)\sigma_m$ for Al-Mg-Y and Al-Fe-Y, respectively.

Since the intermetallic phase (Al$_3$Y) in the Al-Mg-Y alloy does not incorporate the Mg, while that (Al$_{10}$Fe$_2$Y) in the Al-Fe-Y system is composed of all three elements (Fig. 2), the remaining Mg solute atoms within the FCC phase can also provide a solid solution strengthening increment. Based on a prior study of Mg solution hardening in Al [64], an amount of 2 at.% Mg solute can give rise to a yield strength increment of 60 MPa, which is the upper limit of the solute strengthening in the present study since some Mg atoms segregated to grain boundaries.

Amorphous complexions will also contribute a strengthening effect to the present two alloys, as these features have been shown to improve the yield strength of nanocrystalline alloys. For instance, Khalajhedayati et al. [65] performed compression testing on nanocrystalline Cu-Zr micropillars with two different cooling conditions after annealing, and observed that fast quenching to retain the amorphous complexions can increase the yield stress by ~150 MPa as compared to a slow cooling process that leaves ordered grain boundaries. The enhanced yield stress in this case was due only to differences in grain boundary structure since both samples had the same average grain size, alloy composition, and impurity carbide distribution. The strengthening effect of amorphous complexions was also verified by Wardini et al. [66], who showed that the ultimate tensile strength of Cu-3.5 at.% Zr micropillars with an average grain size of ~70 nm increased from 767 MPa to 805 MPa when ordered boundaries were replaced by amorphous complexions. Turlo and Rupert [10] studied the mechanisms of such strengthening with molecular dynamics and uncovered a higher critical stress for dislocation propagation (the rate-limiting mechanism for plasticity in nanocrystalline Cu-Zr) in the presence of amorphous complexions. These authors observed that a higher stress was required for the samples with amorphous complexions to maintain the same dislocation propagation velocity as those with ordered complexions, demonstrating that amorphous complexions restrict dislocation propagation more strongly than ordered grain boundaries. As a whole, these findings suggest that an amorphous grain boundary structure can further increase the material strength in addition to the grain boundary segregation, and therefore previous models such as that by Vo et al. [57] which capture the role of dopant segregation to grain boundaries could be modified by incorporating the contribution due to the amorphous grain boundary complexions. In Ref [15], the activated sintering of Al-Mg-Y was observed to occur at a lower temperature range than that of Al-Fe-Y, suggesting a lower temperature range for the amorphous complexion formation in the former alloy. This, therefore, extends the supercooled liquid window down to lower temperatures in Al-Mg-Y, suggesting an enhanced stability of the complexions in this system. Consequently, the volume fraction of the amorphous complexions may be larger in the Al-Mg-Y alloy than in the Al-Fe-Y alloy, resulting in a more potent strengthening effect in the former system.

**Conclusions**

In the present study, the mechanical behavior of two nanocrystalline Al alloys, Al-Mg-Y and Al-Fe-Y, was investigated using in-situ SEM micropillar compression testing. Both alloys are extremely strong in comparison to other high performance Al alloys, with the maximum yield strength of the Al-Mg-Y alloy being 950 MPa. Two deformation modes were observed in each system, depending on the spatial homogeneity of the intermetallic particles. The following important conclusions are drawn:

1) The nanocrystalline alloys deformed through either stable plastic flow or strain localization into shear bands. Post-mortem SEM and TEM examination revealed dramatic grain coarsening and the same grain orientation within the shear band, pointing to grain boundary-mediated plasticity, specifically grain rotation and/or grain boundary migration, due to the localized deformation.

2) Shear bands were found to occur in regions lacking intermetallic grains, suggesting that a uniform distribution of hard reinforcing particles with an average spacing much smaller than the percolation length of shear localization can effectively prevent localized deformation by frustrating the formation of fully mature shear bands or deflecting incipient ones. The obstructing effect of intermetallic phases on shear band propagation is mainly attributed to the significantly different lattice parameters between the matrix and intermetallic and therefore a higher barrier for activation of grain rotation to facilitate shear band propagation.

3) The exceptional yield strengths of both alloys come from a hierarchical microstructure consisting of grain boundary segregation, amorphous grain boundary complexions with thicknesses of a few nanometers, carbide nanorod precipitates about 20 nm long and 5 nm wide, and submicron-sized intermetallic particles.

4) The higher yield strength of Al-Mg-Y than Al-Fe-Y is mainly attributed to a higher number density of carbide nanorods at grain boundaries, the Mg solute atoms remained in the matrix, and possibly a larger volume fraction of amorphous complexions.

The results of the present study provide insights for designing high-strength nanocrystalline Al alloys. Hierarchical microstructure can enable extremely high strengths, while shear localization can be avoided through the incorporation of a uniform distribution of intermetallic particles.

**Declaration of Competing Interest**

The authors declare that they have no known competing financial interest or personal relationships that could have appeared to influence the work reported in this paper.

**Acknowledgements**

This work was supported by the U.S. Department of Energy, Office of Energy Efficiency and Renewable Energy (EERE), under the Advanced Manufacturing Office Award No. DE-EE0009114. The authors acknowledge the use of facilities and instrumentation at the UC Irvine Materials Research Institute (IMRI), which is supported in part by the National Science Foundation through the UC Irvine Materials Research Science and Engineering Center (DMR-2011967). SEM, FIB, and EDS work was performed using instrumentation funded in part by the National Science Foundation Center for Chemistry at the Space-Time Limit (CHE-0802913).

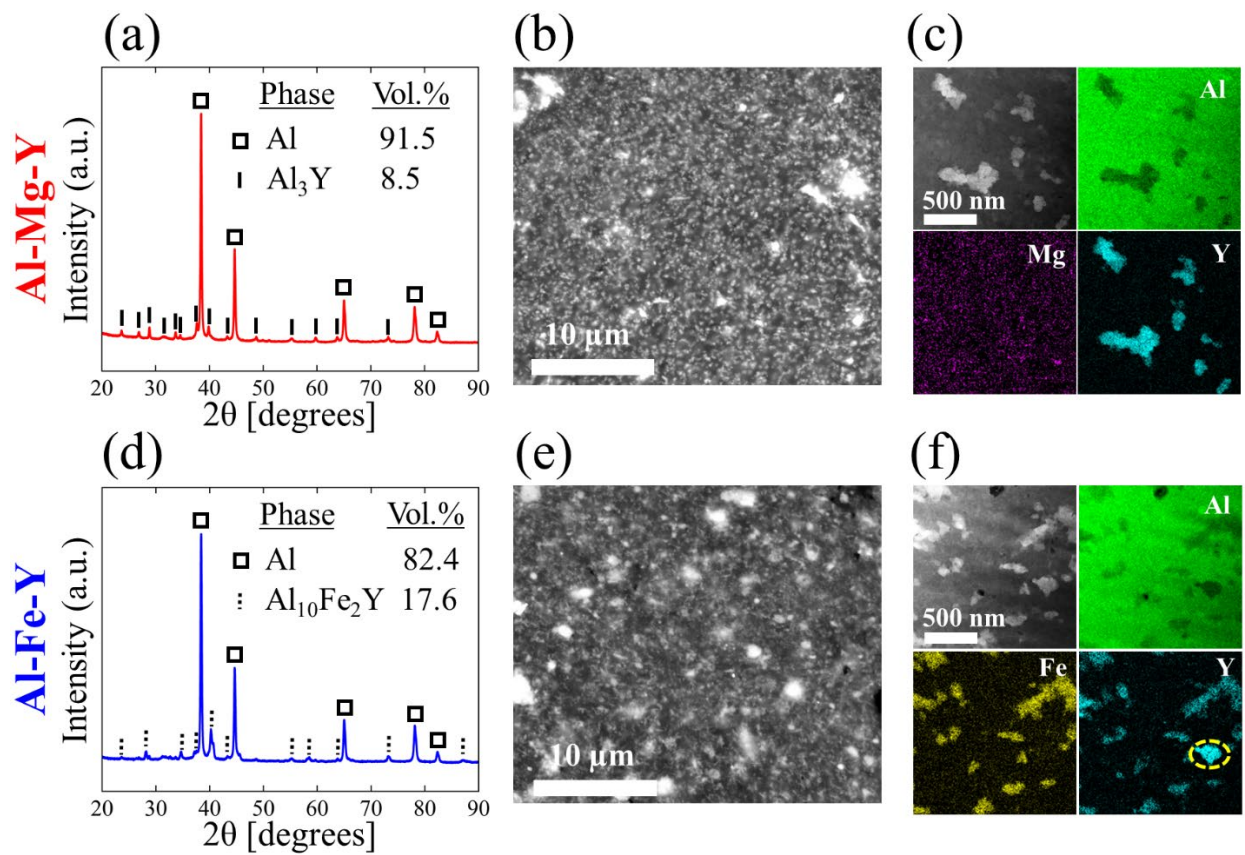

**Figure 1.** Characterization of the intermetallic phases in Al-Mg-Y and Al-Fe-Y. (a) XRD scan, (b) BSE image, and (c) HAADF-STEM and EDS mapping for Al-Mg-Y, revealing the existence of one intermetallic phase, $Al_3Y$. (d) XRD scan, (e) BSE image, and (f) HAADF-STEM and EDS mapping for Al-Fe-Y, where only $Al_{10}Fe_2Y$ phase shows up in the XRD plot but the EDS shows that a few $Al_3Y$ particles also exist (enclosed in a dashed oval). However, the vast majority of the intermetallic particles are $Al_{10}Fe_2Y$ in the Al-Fe-Y alloy.

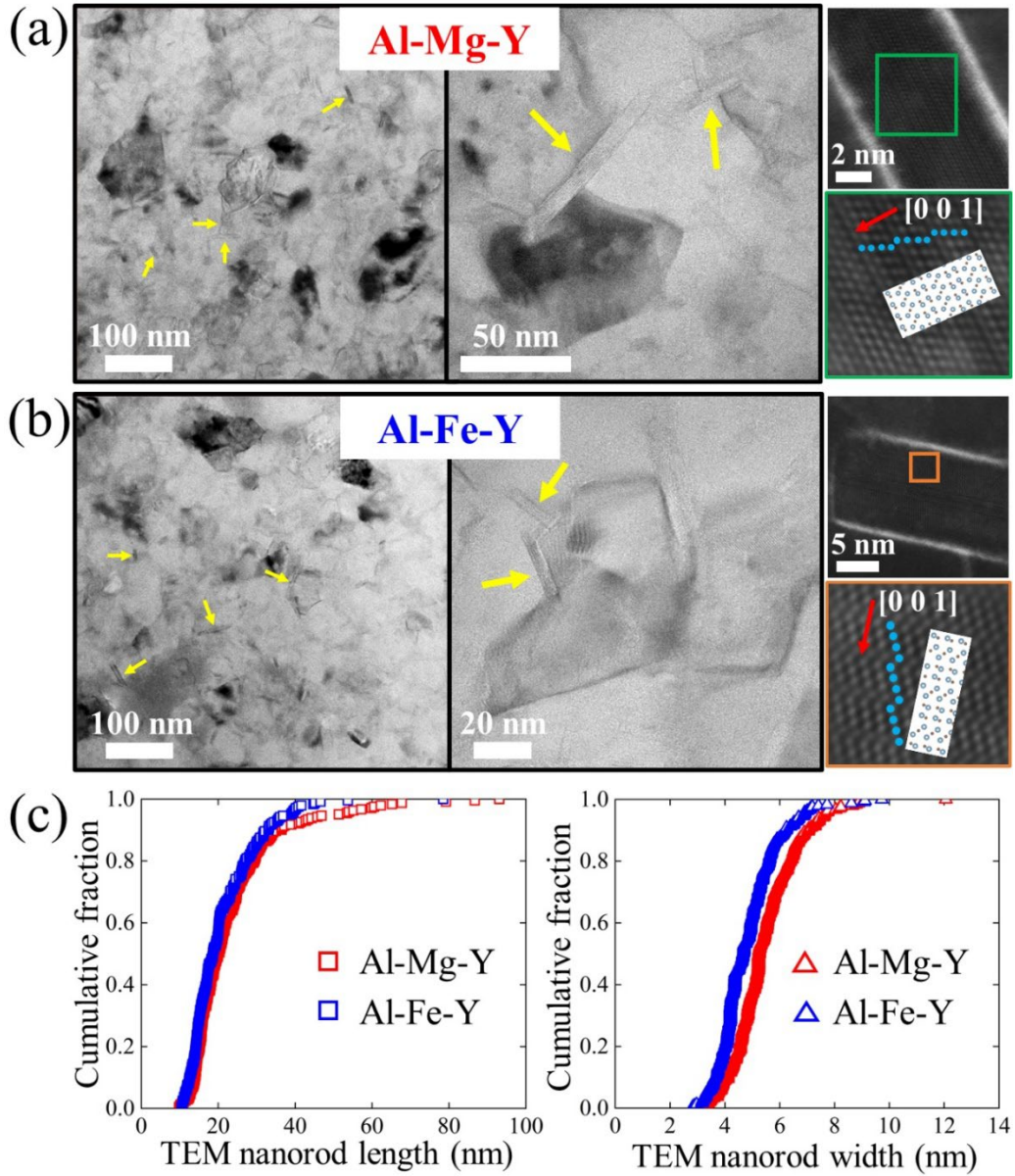

**Figure 2.** Representative bright-field TEM micrographs at different magnifications showing the nanocrystalline grain morphology and location of nanorod precipitates, marked by yellow arrows, for (a) Al-Mg-Y and (b) Al-Fe-Y. The last two panels in (a) and (b) are high-resolution HAADF-STEM micrographs presenting the structure of the nanorod interior in each system, which are consistent with the atomic arrangement of $Al_4C_3$ phase (shown in the inset panels). (c) Cumulative fractions of nanorod length and width measured from over 200 nanorods in each system.

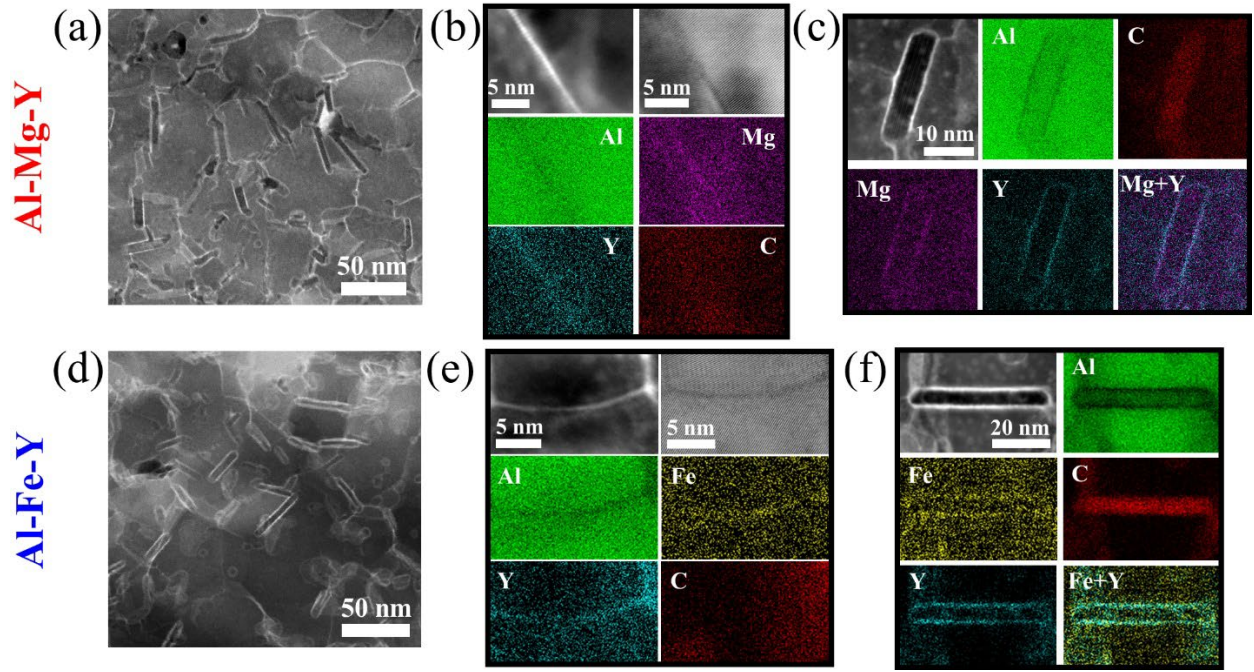

**Figure 3.** HAADF-STEM micrographs and elemental mapping of both grain boundaries and nanorods. (a) and (d) present representative HAADF-STEM micrographs for Al-Mg-Y and Al-Fe-Y, respectively, where grain boundaries and nanorod edges are brighter than the matrix and nanorod interior. (b) and (e) show representative grain boundaries in each system with corresponding elemental mapping, where co-segregation of both dopant elements occurs. (c) and (f) show a representative nanorod in each alloy and the distribution of elements in the same region, with co-segregation again observed.

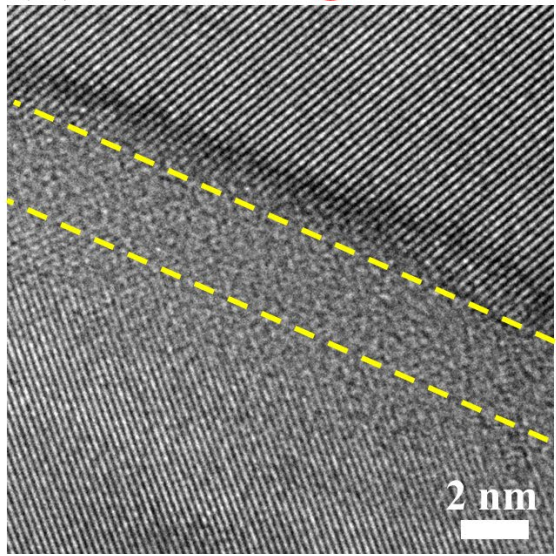 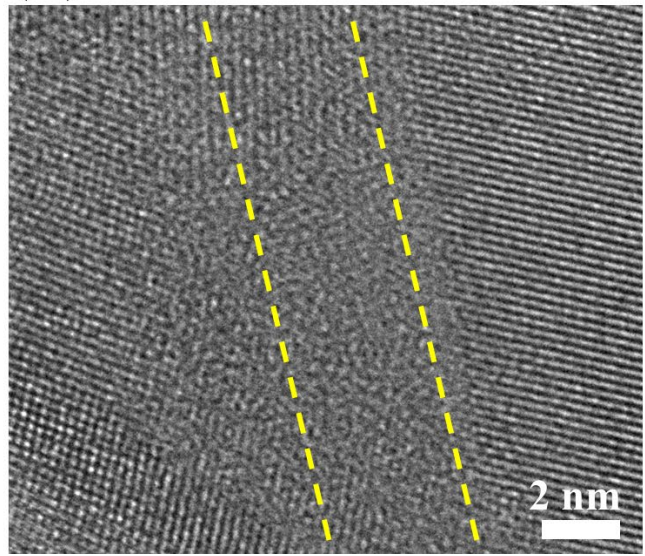

**Figure 4.** High resolution TEM of representative amorphous complexions observed in (a) Al-Mg-Y and (b) Al-Fe-Y.

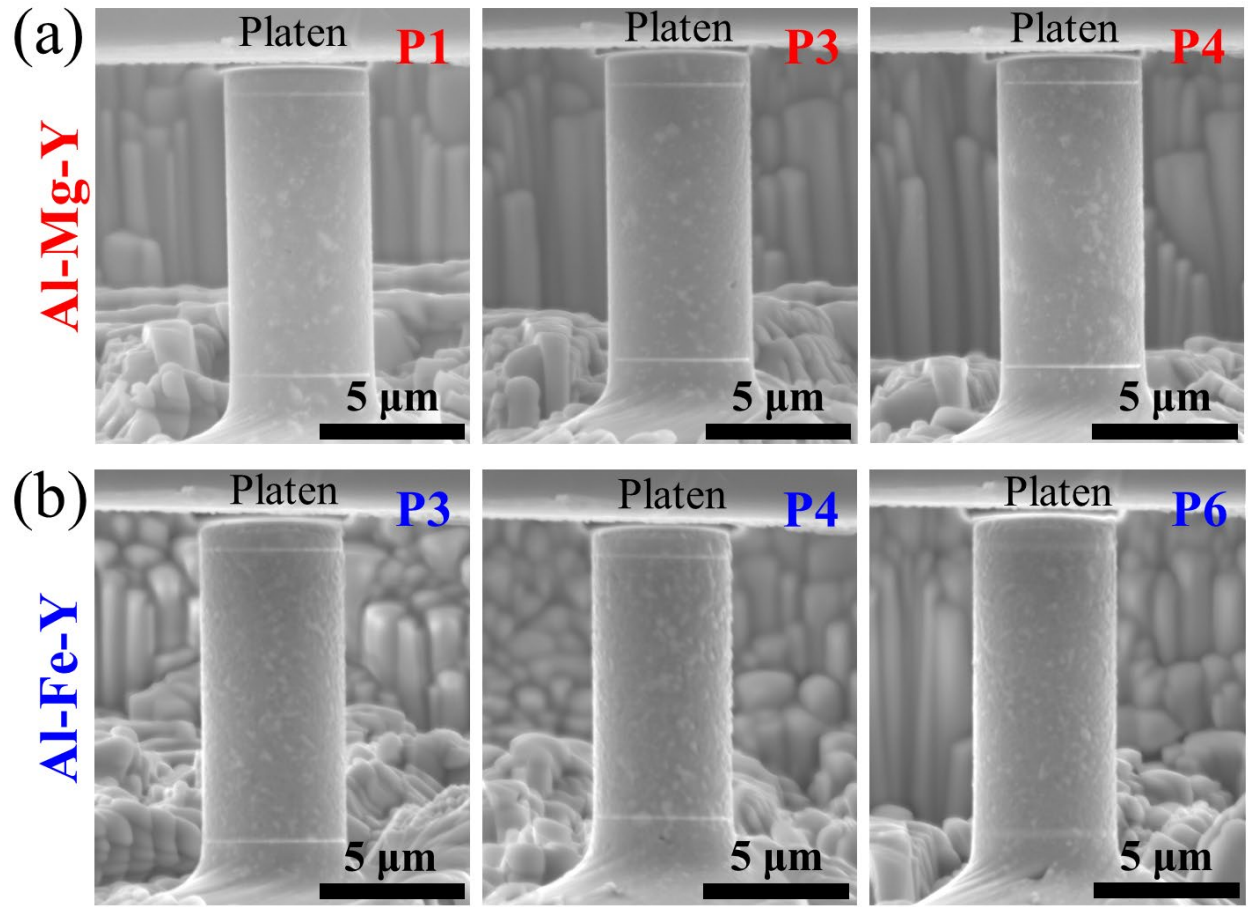

**Figure 5.** SEM micrographs of three representative pillars before compression testing for (a) Al-Mg-Y and (b) Al-Fe-Y. Taper-free pillars are achieved due to the use of a lathe milling method, and all pillars are ~5 μm in diameter and ~10-11 μm in height to ensure an aspect ratio of approximately 2 to prevent plastic buckling. The platen is ~0.2 μm above each pillar in these images.

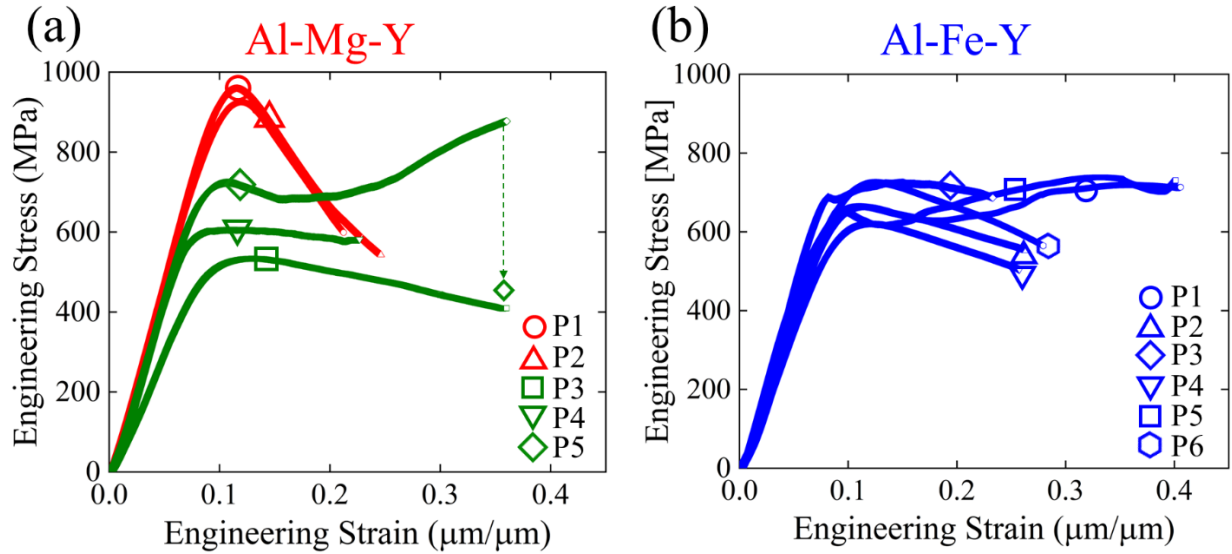

**Figure 6.** Engineering stress-strain curves of micropillar compression tests for (a) Al-Mg-Y and (b) Al-Fe-Y alloys. For the Al-Mg-Y system, the stress of Pillar 5 showed an increasing trend after yielding, which was due to increasing contact area at the top surface during deformation rather than a strain hardening effect. The true stress of the last data point is corrected based on post mortem imaging, which was much lower than the corresponding engineering value. For the Al-Fe-Y alloy, some pillars also exhibited an increasing stress after yielding, which was also due to the increasing top contact area during deformation.

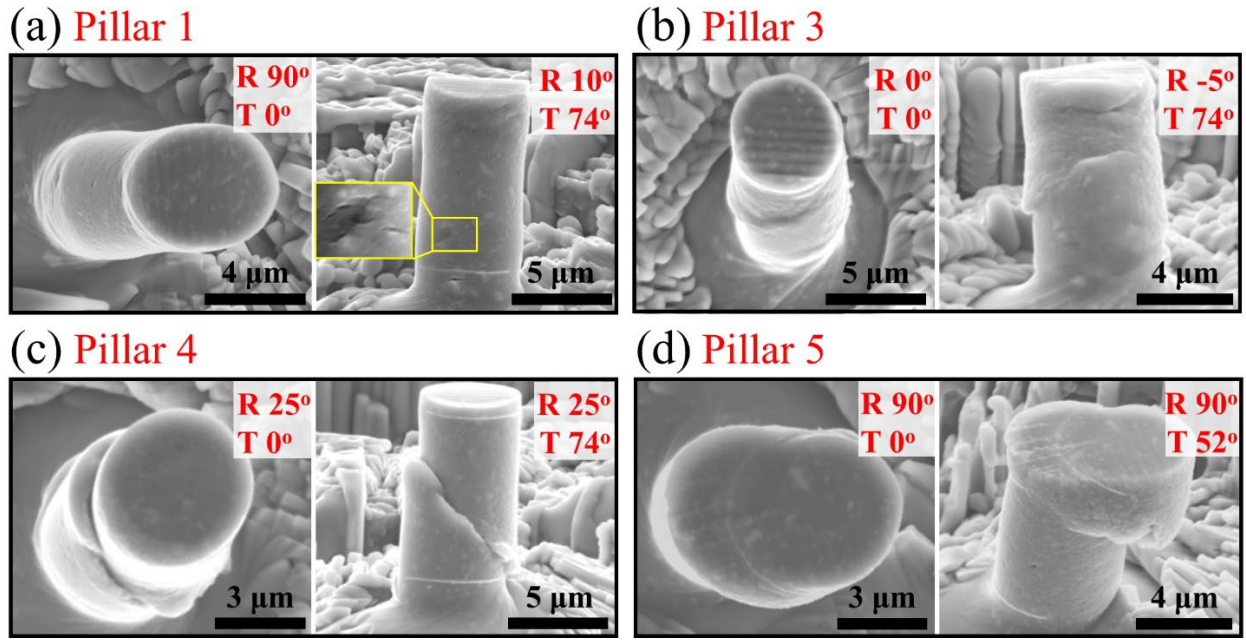

**Figure 7.** SEM micrographs at different rotation (R) and tilt (T) angles of four deformed pillars for the Al-Mg-Y alloy. For each pillar, the first image is a top-down view and the second one is a front view. Two deformation modes were observed, as Pillar 1 experienced steady plastic flow with homogeneous deformation (Pillar 2 also deformed in this fashion, but is not shown here), while Pillars 3-5 failed through strain localization within shear bands.

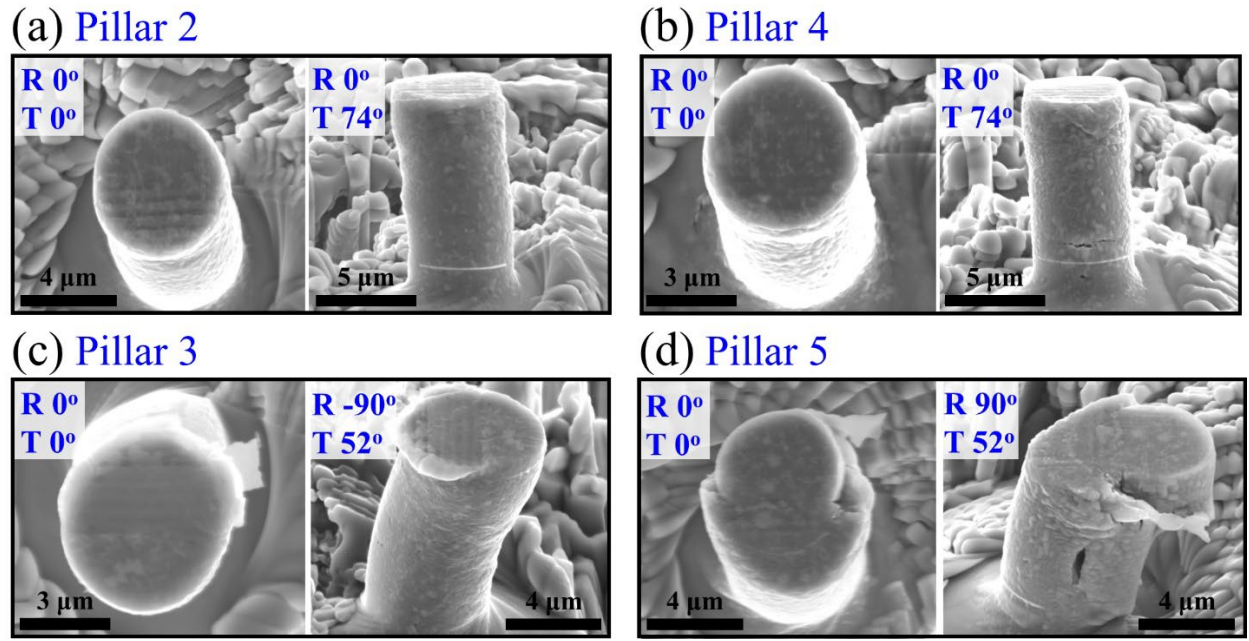

**Figure 8.** SEM micrographs at different rotation (R) and tilt (T) angles of four deformed pillars for Al-Fe-Y. For each pillar, the first image is a top-down view and the second one is a front view. Two deformation modes were observed, steady plastic flow (Pillars 2 and 4) and shear localization (Pillars 3 and 5), similar to those observed in Al-Mg-Y.

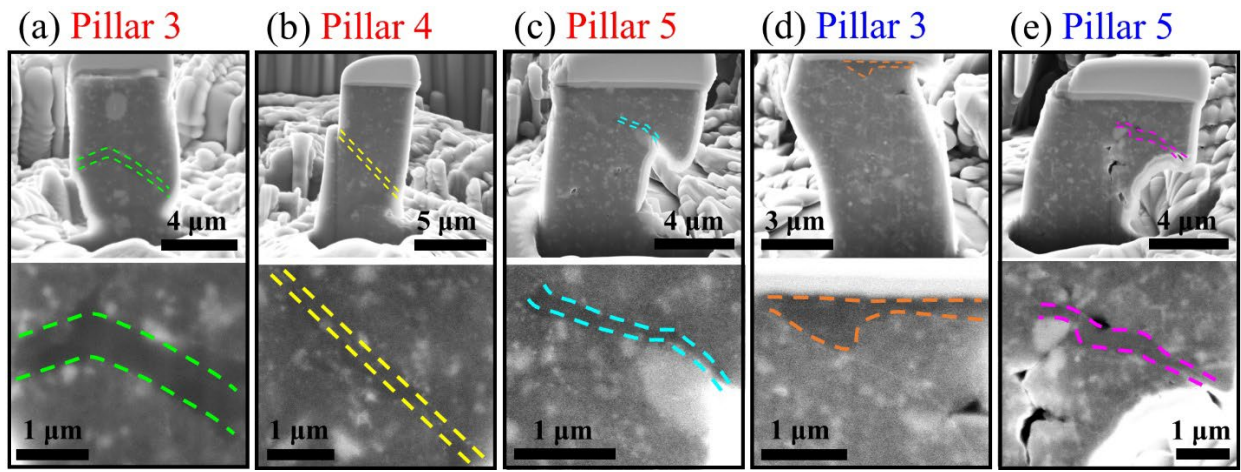

**Figure 9.** Cross-sectional SEM micrographs of deformed pillars that failed through shear localization for (a)-(c) Al-Mg-Y, (d)-(e) Al-Fe-Y. In each pillar, a darker pathway free of intermetallic particles was observed and is denoted in these images by dashed lines. Moreover, the location of the pathway was consistent with that of the shear band in all of the pillars.

(a) Pillar 1 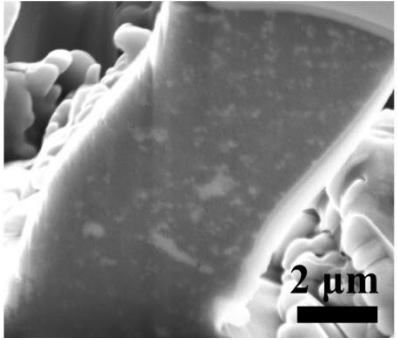 (b) Pillar 2 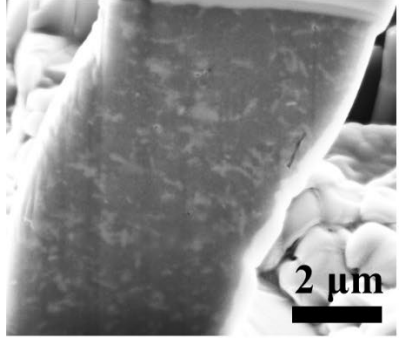 (c) Pillar 4 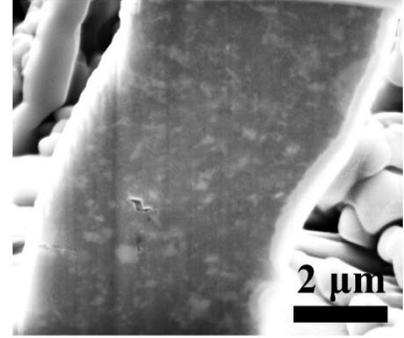

**Figure 10.** Cross-sectional SEM micrographs of deformed pillars experienced homogeneous deformation for (a) Al-Mg-Y and (b)-(c) Al-Fe-Y. No darker pathway free of intermetallic phases was observed.

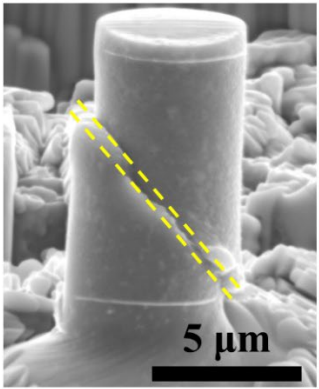
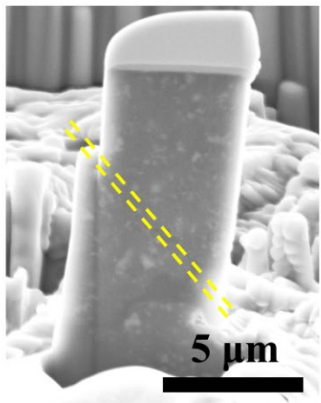
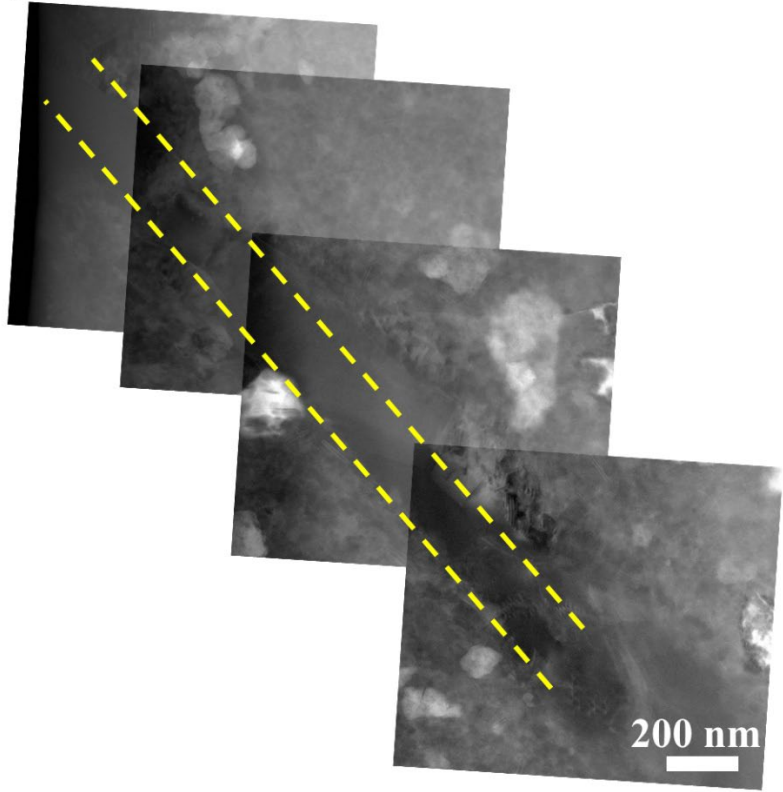

**Figure 11.** (a) SEM and (b) HAADF-STEM micrographs corresponding to the shear band in Pillar 4 of the Al-Mg-Y system, confirming that no intermetallic particles exist within the shear band.

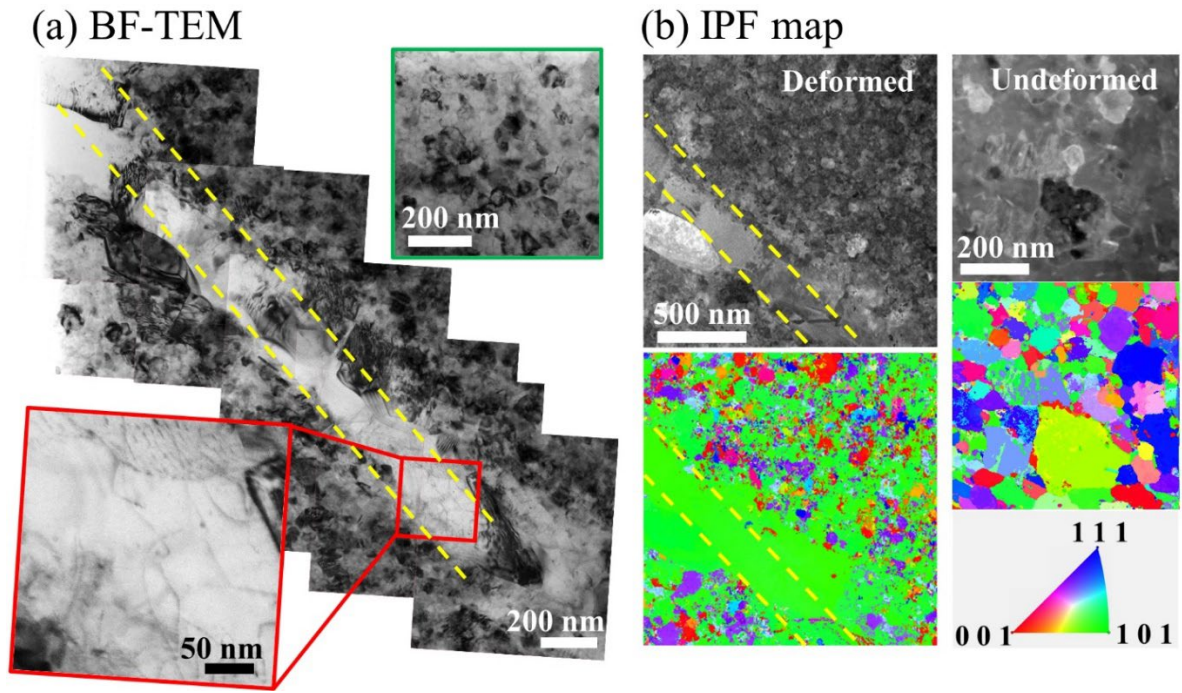

**Figure 12.** (a) BF-TEM micrographs of the deformed Pillar 4 for Al-Mg-Y, where the dominant shear band is enclosed in outlined by dashed lines. The grains within the shear band significantly coarsen and intragranular dislocation accumulation was observed in the coarsened grains, as shown in the magnified view (red outline). One micrograph corresponding to an area far away from the shear band is also presented (green outline), where the grain sizes are well below 100 nm. (b) Inverse pole figure (IPF) maps of both the shear band region and an undeformed region of the sample obtained from ASTAR automated crystal orientation map. The grains within the shear band have the same orientation, pointing to grain rotation and/or grain boundary migration within the shear band, while no preferred texture exists in the sample without deformation.

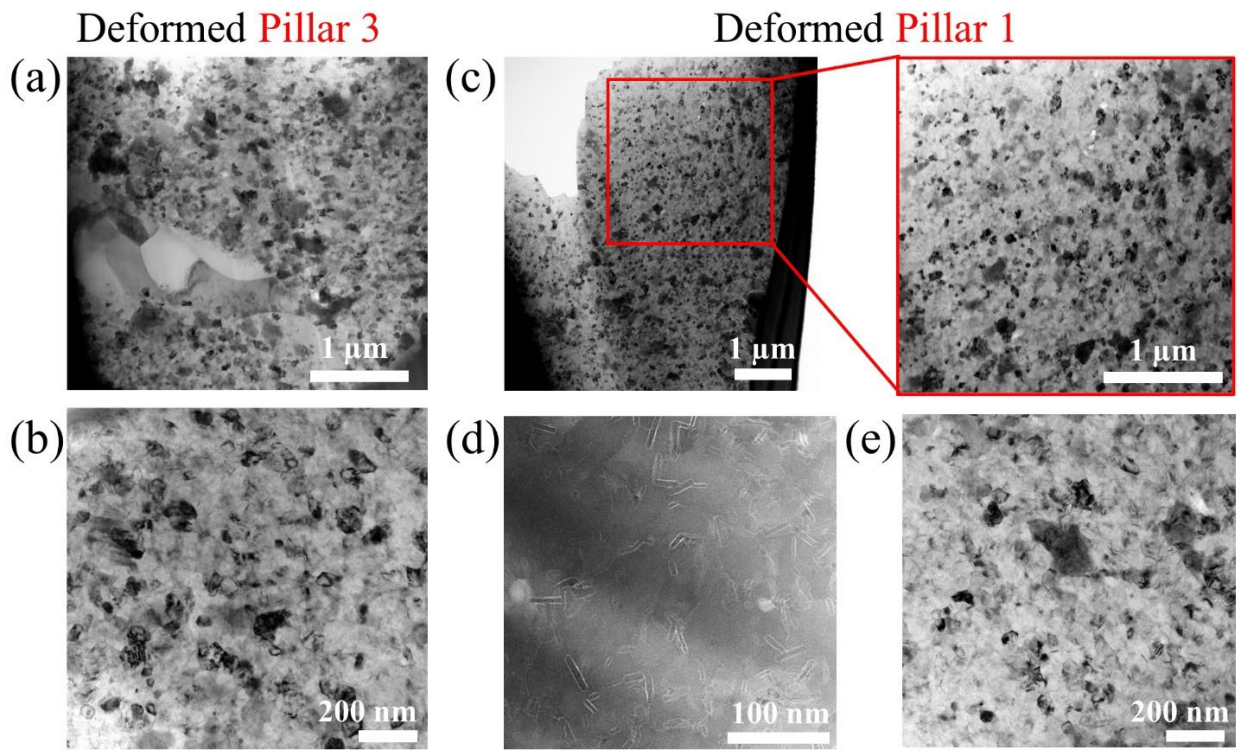

**Figure 13.** (a) Low-magnification BF-STEM micrograph of the deformed Pillar 3 for Al-Mg-Y showing significant grain coarsening in the middle region, where the dominant shear band was located. (b) BF-TEM micrograph presenting a magnified view of the microstructure away from the localized deformation, where the grains remain below 100 nm. (c) Low-magnification BF-STEM micrograph along with one zoomed-in view of a selected region of the deformed Pillar 1 for Al-Mg-Y that experienced stable plastic flow. The microstructure only consists of nanosized grains. (d) HAADF-STEM micrograph presenting the size, morphology, and distribution of nanorod carbides in the deformed Pillar 1, all of which are consistent with those in the undeformed condition. (e) An additional BF-TEM micrograph of one representative region in the deformed Pillar 1, which is very similar to those away from the localized deformation in the deformed Pillars 3 and 4.

**Table I.** Hardness values obtained from nanoindentation tests [15] and yield strengths of micropillars without shear localization.

| Alloy composition (at.%) | Nanoindentation hardness (GPa) | Micropillar compression yield strength (MPa) | Hardness/Strength Ratio |
|---|---|---|---|
| Al-2Mg-2Y | 2.77 ± 0.12 | 920 ± 42 | 3.01 |
| Al-2Fe-2Y | 2.18 ± 0.15 | 613 ± 58 | 3.56 |